\documentclass[10pt,aps,pra,notitlepage,nofootinbib,twocolumn,superscriptaddress,noeprint]{revtex4-2}
\usepackage[utf8]{inputenc}
\usepackage[english]{babel}
\usepackage[T1]{fontenc}
\usepackage{amsmath,amsfonts,amssymb,amsthm,bm,bbm,bbold}
\usepackage{graphicx}
\usepackage{mathtools}
\usepackage{physics}
\usepackage{makecell}
\usepackage{textcomp}
\usepackage{microtype}
\usepackage[dvipsnames]{xcolor}
\usepackage{dsfont}
\usepackage{booktabs}
\usepackage[breaklinks=true,colorlinks=true,linkcolor=teal,urlcolor=teal,citecolor=teal]{hyperref}
\usepackage{orcidlink}
\begin{document}
\title{Charging of a Quantum Battery by a Single-Photon Quantum Pulse}
	
\author{Elnaz Darsheshdar\,\orcidlink{0000-0001-6341-7151}}
\thanks{darsheshdare@gmail.com}
\affiliation{Department of Physics and Astronomy, University of Exeter, Exeter, EX4 4QL, United Kingdom}
\affiliation{Dipartimento di Fisica, Universit\`a di Napoli Federico II, Complesso Universitario di Monte S. Angelo, Via Cintia, 80126 Napoli, Italy}

\author{Seyed Mostafa Moniri\,\orcidlink{0000-0003-1738-4429
}}
\thanks{s.m.moniri@gmail.com}
\affiliation{Basic Sciences Group, Golpayegan College of Engineering, Isfahan University of Technology, Golpayegan 87717-67498, Iran}

\begin{abstract}

We study a minimal model for charging a quantum battery consisting of a two-level system (TLS) acting as a charger, coupled to a harmonic oscillator that serves as the quantum battery. A single-photon quantum pulse of light excites the TLS, which subsequently transfers its excitation to the isolated battery. The TLS may also decay into the electromagnetic environment. 
We obtain analytical solutions for the dynamics of the battery and determine the optimal pulse shape that maximizes the stored energy. The optimal pulse saturates a universal bound for the stored energy, determined by the TLS decay rates into the pulse and the environment. Furthermore, we derive the minimum charging time and establish a quantum speed limit at the exceptional point, where a critical transition occurs in the system’s dynamics. We also present analytical expressions for the charging power and investigate the pulse duration that maximizes it.

\end{abstract}

\maketitle

\section{Introduction}\label{sec:int}

Quantum batteries have emerged recently, aiming to use quantum resources to increase energy storage and charging performance~\cite{Ferraro2026}. The model and concept of quantum batteries were introduced by Alicki and Fannes~\cite{Alicki2013}. They defined quantum batteries as finite-dimensional quantum systems effective for storing extractable work, i.e., ergotropy~\cite{Nieuwenhuizen2004}, as a key figure of merit in charging protocols. Further investigations showed that collective charging and operation enable quantum speed-up in the power of charging compared to local ones, thus highlighting the effect of entanglement and many-body effects~\cite{Binder2015,Campaioli2017,Huber2015}. Further works expanded these ideas to realistic scenarios in addition to studying finite-time thermodynamics, power--efficiency trade-offs, and the impact of system size and interaction topology~\cite{Campaioli2024,Andolina2019}. In recent years, models that incorporate open-system dynamics, dissipation, and experimentally accessible interactions, including charging driven by quantum fields and photonic environments, have been studied more exactly~\cite{Ferraro2018,Gherardini2020}.

Light–matter interaction is an attractive setup for quantum energy storage \cite{Gorshkov2007,Keller2004,Reiserer2015}, and photon pulses that match the response of a quantum emitter
have been studied extensively \cite{Das2025}. In recent works, the charging process of a minimal quantum battery under classical light excitation has been investigated ~\cite{Downing2024,Downing2024hyperbolic,Downing2024two,Downing2025energy}. Despite these advances, a systematic investigation of quantum battery charging in realistic open-system conditions, such as charging by pulsed few-photon quantum light, remains incomplete. This requires the use of methods from quantum thermodynamics, specifically those involving light-matter interactions, to characterize optimal charging strategies that extend beyond idealized limits. In modern quantum-optical experiments \cite{Karpinski2017,Karpinski2021}, single-photon pulse shapes can be engineered, making them suitable candidates for optimal charging of a quantum-battery device. 

In this work, we provide an experimentally applicable model for quantum battery charging, where energy is transferred from a travelling single-photon pulse to a harmonic oscillator, mediated by a TLS. We present fully general expressions describing the energetic and power performance of the battery, valid for arbitrary pulse shapes. This allows us to obtain optimal pulse shapes that maximize energy storage and ergotropy. We also present a minimization of the charging time, which relates our theoretical model to single-photon experiments and enables the design of realistic charging protocols in noisy quantum optical settings.
Our main results are presented in Figs \ref{fig:8}, \ref{fig:7} and Eq.~\eqref{xi_opt}. They are obtained by optimizing the ergotropy of the quantum battery. Importantly, the proposed charging mechanism is directly compatible with existing experimental platforms in both optical and microwave quantum technologies.
\paragraph*{Experimental platforms.}
The model presented in this work can be experimentally applied in both optical and microwave setups, where a propagating light or microwave pulse excites a TLS or a superconducting qubit (e.g., an atom, quantum dot, or transmon) coupled to a long-lived optical cavity mode or a high-$Q$ microwave resonator that acts as the quantum battery. 
In both optical and microwave setups, the decay rates in the guided mode and the environment correspond to $\Gamma$ and $\Gamma_\perp$ in our model. Existing pulse-shaping techniques enable the engineered temporal envelopes required to approximate the optimal single-photon pulses identified in this work.

\section{THEORETICAL FRAMEWORK}\label{sec:model}
\begin{figure}[ht]
	\centering
	\includegraphics[width=0.9\linewidth]{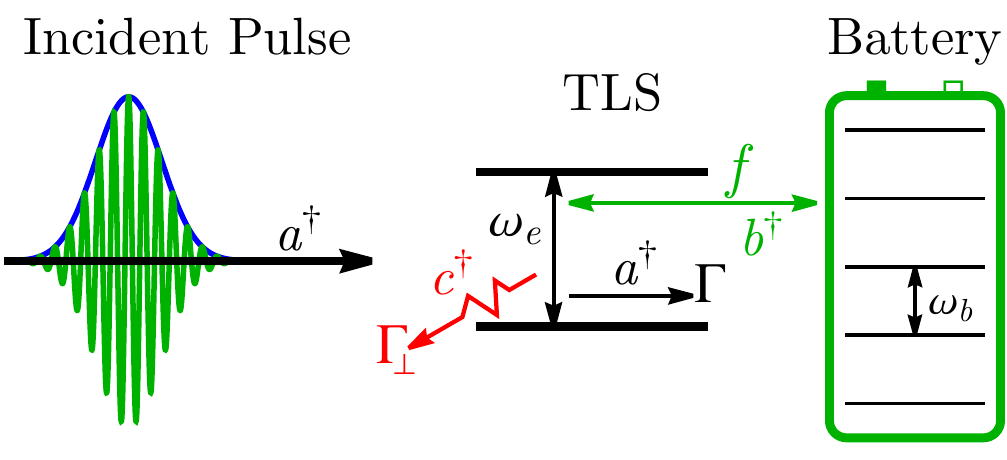}
	\caption{Illustration (not to scale) of charging a harmonic oscillator as a battery with a single-photon pulse. $\Gamma$ ($\Gamma_\perp$) captures the interaction strength with the pulse (environment).  The parameter $\omega_b$ denotes the frequency of the harmonic oscillator.  The parameter $f$ is the coherent coupling strength governing the exchange of single excitations between the TLS and the harmonic oscillator.}
	\label{fig:qb_schematic}
\end{figure}

We investigate a minimal quantum charger--battery architecture consisting of a TLS acting as a charger and a harmonic oscillator (HO) acting as the battery. A single-photon traveling pulse excites the system by passing through the TLS, which then transfers its excitation to the HO through a coupling $f$. The TLS may also decay into the pulse mode at a rate $\Gamma$ or into the environment at a rate $\Gamma_\perp$ \cite{Ko2022,Baragiola2014open,Wang2011,Silberfarb2004}. A schematic of the model is shown in Fig.~\ref{fig:qb_schematic}. This setup is the same as that considered in many quantum spectroscopic schemes \cite{Albarelli2023,Darsheshdar2024,Khan2024,Khan2025}. However, instead of detecting the light after its interaction with the TLS, we store its energy in the harmonic oscillator as a quantum battery.
    
The quantum battery here is a finite-dimensional system that is in the ground state at $t=0$. In this work, the HO battery is effectively restricted to the single-excitation subspace $\{|0\rangle,|1\rangle\}$ consistent with the single-photon drive, so non-passivity is related to the excited-state population. The free  Hamiltonian of the system is defined by~\cite{Blow1990a} (throughout this work, we set $\hbar=1$),
\begin{align}\label{H0}
	H_0 & =\omega_e\,\sigma_z+\omega_b\,b^\dagger b \\ \nonumber
    &+\!\int_0^\infty d\omega\,\omega\,a^\dagger(\omega)a(\omega)
	+\!\int_0^\infty d\omega\,\omega\,c^\dagger(\omega)c(\omega).
\end{align}
Here, $\omega_e$ is the transition frequency of the TLS between its ground state $\ket{g}$ and excited state $\ket{e}$, 
$\omega_b$ is the frequency of the HO, $b$ and $b^\dagger$ are its annihilation and creation operators satisfying $[b, b^\dagger]=1$, 
and $\sigma_z = \ket{e}\bra{e}$ is the Pauli-$z$ operator for the TLS \cite{Scully1997}.

The third term in Eq.~\eqref{H0} corresponds to a traveling pulse of quantum light, modeled as a continuum of white-noise bosonic modes $a(\omega)$ that satisfy 
$[a(\omega), a^\dagger(\omega')] = \delta(\omega - \omega')$. 
In addition, TLS in free space interacts with an infinite set of all other electromagnetic environmental modes of the remaining spatial and polarization degrees of freedom beyond those of the pulse. 
Thus, we introduce an additional continuum of bosonic modes $c(\omega)$, whose free Hamiltonian is given by the final term in Eq.~\eqref{H0}.

Both the pulse and the environmental fields couple to the TLS through the standard dipole interaction 
$-\boldsymbol{\mu} \cdot \mathbf{E}$, 
where $\boldsymbol{\mu}$ is the TLS dipole operator and $\mathbf{E}$ is the total electric field of the two fields.
The electric field operator, corresponding to a well-defined direction of propagation, is 
\begin{equation}\label{eq:Electric_field}
\mathbf{E}(t) = i  \bm{\epsilon} \mathcal{A}(\bar{\omega})  \int_{-\infty}^{\infty} \frac{d\omega}{\sqrt{2\pi}} \; a(\omega) e^{-i \omega t},
\end{equation}
where $\bm{\epsilon}$ represents the unit polarization vector. The function $\mathcal{A}(\bar{\omega})$ is defined as $\sqrt{{\bar{\omega}}/{2 \epsilon_{0}c A }}$, where $\bar{\omega}$ denotes the central frequency of the pulse, and $A$ refers to the transverse quantization area.
The integral in Eq.(\ref{eq:Electric_field}) spans the entire $\omega$-axis, incorporating the slowly-varying envelope approximation~\cite{Albarelli2023,Darsheshdar2024,Schlawin2017,Deutsch1991} that applies to paraxial pulses considered here.

The TLS also couples coherently to the HO, which acts as the quantum battery by the Pauli $x$ operator $\sigma_x=\sigma^+ + \sigma^-$, with $\sigma^+=|e\rangle\langle g|$ and  $\sigma^-=|g\rangle\langle e|$. This interaction is described by the exchange of single excitations with coupling strength $f$. This coherent link governs the energy storage in the battery.

In the picture generated by the unitary transformation $e^{-\mathrm{i} H_0 t}$ \cite{DeBernardis2018}, and using the slowly varying envelope~\cite{Albarelli2023,Darsheshdar2024,Schlawin2017,Deutsch1991}, rotating wave and dipole approximations~\cite{Mandel1995}, the Hamiltonian governing the dynamics of the TLS, the HO, the probe field, and the electromagnetic environment is given by~\cite{Scully1997,Ko2022}
\begin{align}\label{Ham_int}
	H_I(t) &= f\, \left( \sigma^+ b\, e^{i(\omega_e - \omega_b)t} + \sigma^- b^\dagger\, e^{-i(\omega_e - \omega_b)t} \right) \\ \nonumber
	&- i \left( \sqrt{\Gamma}  \sigma^+  a(t) - \text{h.c.} \right) - i \left( \sqrt{\Gamma_\perp}  \sigma^+  c(t) - \text{h.c.} \right).
\end{align}

In Eq. \eqref{Ham_int} we have introduced the so-called ``quantum white-noise'' operators \cite{Ko2022},
\begin{equation}
    \label{eq:white_noise_a}
    a(t) = \int_{-\infty}^{\infty} \frac{d\omega}{\sqrt{2\pi}} \; a(\omega) e^{-i (\omega-\bar{\omega}) t},
\end{equation}
satisfying $[a(t),a^\dag(t')] = \delta(t-t')$ as well as the coupling constant $\Gamma$ which is proportional to the
square of the dipole moment.
This Hamiltonian provides the foundation for analyzing the charging dynamics of the quantum battery driven by a single-photon pulse. 
The effects of environmental loss, i.e., $\Gamma_\perp$, and pulse-mediated decay, $\Gamma$, are studied separately. The loss to the environment can be tuned in experimental setups, ranging from zero in the case of an atom interacting with a field propagating in a unidirectional waveguide, to nonzero values representing light emitted in the direction opposite to the pulse propagation~\cite{Molmer2024}, or in free-space scenarios within the paraxial approximation~\cite{Ko2022,Silberfarb2004,Cook2023}.

We focus on a single-photon pulse~\cite{Blow1990a,Fabre2020modes} 
\begin{equation}
\ket{1_{\xi}} = \int_{-\infty}^\infty d\tau ~\xi(\tau) a^\dag (\tau)  \ket{0}_P,
\end{equation}
where $\xi(\tau)$ is the normalized temporal profile of the pulse. Assuming that the TLS and HO start in their ground states, i.e., $\ket{g}$ and $\ket{0}_B$, and that the environment is in the vacuum state $\ket{0}_E$, the joint quantum state of the system after evolution under the Hamiltonian in Eq.~\eqref{Ham_int} is
\begin{align}\label{ansatz2}
	\ket{\Psi(t)} &= \alpha_0^b(t)\ket{e}\ket{0}_B\ket{0}_{P}\ket{0}_{E} \\ \nonumber
	&+ \alpha_1^b(t)\ket{g}\ket{1}_B\ket{0}_{P}\ket{0}_{E}\\ \nonumber
	&+ \ket{g}\ket{0}_B\int^\infty_{-\infty} d\tau \chi_{0,P}(t,\tau) a^\dagger(\tau)\ket{0}_{P}\ket{0}_{E} \\ \nonumber
    &+ \ket{g}\ket{0}_B\int^\infty_{-\infty} d\tau \chi_{0,E}(t,\tau) c^\dagger(\tau)\ket{0}_{P}\ket{0}_{E}.
\end{align}

Here, $\alpha_0^b(t)$ and $\alpha_1^b(t)$ denote the amplitudes of the battery ground and excited states, respectively, 
while $\chi_{0,P}(t,\tau)$ and $\chi_{0,E}(t,\tau)$ describe excitations emitted respectively 
into the pulse and into the environment.
 The total excitation number in the system is conserved to one, which restricts our ansatz to exclude components with more than one excitation in all degrees of freedom.  In this system, if $|\alpha_1^b(t)|^2 > 1/2$, the battery is active and has nonzero ergotropy.  The stored energy and the ergotropy are
\begin{align}
	&E(t) = \omega_b|\alpha_1^b(t)|^2 , \\ 
	&W_{\mathrm{erg}}(t) = 
    \begin{cases}
		0, & |\alpha_1^b(t)|^2 \le 1/2, \\[4pt]
		\omega_b \left(2|\alpha_1^b(t)|^2 - 1 \right), & |\alpha_1^b(t)|^2 > 1/2 .
	\end{cases} \label{erg}
\end{align}
 
 These are analogous to previous QB models based on coupled harmonic oscillators~\cite{Downing2024,Downing2024hyperbolic,Downing2025energy}, but here charging is mediated by a TLS and driven by a single-photon pulse, thereby allowing the temporal shape of the input pulse to be engineered to optimize performance.
 
We solve the Schr{\"o}dinger equation of $\frac{\partial}{\partial t}\ket{\Psi(t)} = -i H^I(t)\ket{\Psi(t)}$ for the interaction picture Hamiltonian \eqref{Ham_int} assuming the initial state
\begin{equation}
|\Psi(-\infty)\rangle = |g\rangle |0\rangle_B \otimes \Big( \int d\tau \, \xi(\tau) \, a^\dagger(\tau) \Big)|0\rangle_P \otimes |0\rangle_E,
\end{equation}
and $\omega_e \simeq \omega_b$. Depending on the dynamical regime characterized by $\kappa = \gamma^2 - 4f^2$, 
 $\alpha_1^b(t)$ is as follows (see Appendix~\ref{App:A}):

\begin{align}\label{eq:alphag}
	&\alpha_1^b(t)= \\ \nonumber
	&\begin{cases}
		if\frac{\sqrt{\Gamma}}{{\Omega}}\displaystyle\int_{t_0}^t e^{-\gamma (t-\tau)/2} \sin\!\Big(\Omega\,(t-\tau)\Big) \xi(\tau)\,d\tau,&\!\kappa<0,\\[12pt]
		if\sqrt{\Gamma}\displaystyle\int_{t_0}^t (t-\tau)e^{-\gamma (t-\tau)/2}\,\xi(\tau)\,d\tau,&\!\kappa=0,\\[12pt]
		2if \sqrt{\frac{\Gamma}{\kappa}} \displaystyle\int_{t_0}^{t} e^{-\gamma (t-\tau)/2} \sinh\!\Big(\tfrac{\sqrt{\kappa}(t-\tau)}{2}\Big) \xi(\tau)\,d\tau,&\!\kappa>0.
	\end{cases}
\end{align}
where $\gamma=(\Gamma+\Gamma_\perp)/2$, $\Omega=\sqrt{f^2-\gamma^2/4}$ and $t_0$ is the starting time of the pulse. We define the exceptional point (EP) at $\kappa=0$ (equivalently $f=f_{\mathrm{EP}}\equiv \gamma/2$), which separates the underdamped ($\kappa<0$) and overdamped ($\kappa>0$) regimes and corresponds to the fastest non-oscillatory charging of the battery. Using an idealized delta pulse of $\xi(t) = \delta(t-t_0)/\sqrt{\Gamma}$ reproduces the results of our previous work~\cite{Downing2024} (plots not shown here), demonstrating the consistency of the model. In Appendix~\ref{App:C}, we provide analytical expressions for the charging dynamics of three pulse shapes, while Appendix~\ref{sec:gauss} discusses the Gaussian case with illustrative plots.

\section{Optimal Battery Efficiency}\label{sec:opt_pulse}

\subsection{Optimal Pulse Shape}

So far, we have analyzed the charging dynamics for an arbitrary input pulse shape of $\xi(t)$. Using the expressions of $|\alpha_1^b(t)|^2$, we reply to the more general question of identifying the optimal single photon pulse shape that maximizes the excitation of the battery mode. Specifically, we aim to determine the temporal profile $\xi_{\mathrm{opt}}(t)$ that yields the highest occupation $|\alpha_1^b(t)|^2$ at the end of the charging process, given fixed decay rates and coupling strength. This problem can be formulated as a linear functional optimization over the input field, resulting in an integral kernel equation whose solution yields the optimal pulse shape. We define the kernel $K_t(\tau)$ as
\begin{equation}
	K_t(\tau) = i f \sqrt{\Gamma}\, G(t-\tau)\,\Theta(t-\tau).
\end{equation}
where $G(t)$ is the Green's function that represents the response of the TLS--HO system defined in Eq.~(\ref{eq:GreenFunction}) and $\Theta(t)$ is the Heaviside function. We can write $\alpha_1^b(t)$ of Eq.~\eqref{eq:alphag0} in the standard convolution form,
\begin{equation}
	\alpha_1^b(t)= \int_{-\infty}^{\infty} K_t(\tau)\,\xi(\tau)\,d\tau = (K_t*\xi)(t),
\end{equation}
where $\xi(\tau)$ is the temporal profile of the incident photon pulse and $*$ is the convolution operator. The kernel $K_t(\tau)$ is a causal response function, vanishing whenever $\tau>t$. This means that the atomic excitation amplitude at time $t$ will not depend on the portions of the pulse that have not yet arrived. Thus, $\alpha_1^b(t)$ takes the form of an inner product between the causal kernel $K_t$ and the input $\xi$, which enables us to use the Cauchy--Schwarz inequality to determine the optimal input pulse shape. Thus we have
\begin{equation}
	|\alpha_1^b(t)|^2 \leq \|K_t\|^2\,\|\xi\|^2.
\end{equation}
The input pulse is normalized  i. e., $\|\xi\|^2=\int |\xi(\tau)|^2\,d\tau=1$, thus the maximum possible excitation probability can be obtained
\begin{equation}
	|\alpha_1^b(t)|^2 \leq \|K_t\|^2.
\end{equation}
The equality is fulfilled only if the two functions are linearly dependent, thus we have $
	\xi(\tau) = c\,K_t(\tau),$  $c$ fixed by the normalization condition. The optimal pulse is
\begin{equation}\label{eq:xiopt0}
	\xi_{\mathrm{opt}}(\tau) = \frac{K_t(\tau)}{\|K_t\|}
	= \frac{f \sqrt{\Gamma}}{\|K_t\|}\,G(t-\tau)\,\Theta(t-\tau),
\end{equation}
 where a global phase does not affect the probability or expectation value and has been absorbed into unity. The $L^2$ norm of the kernel is
\begin{align}
	\|K_t\|^2 
	&= \int_{-\infty}^{\infty} |K_t(\tau)|^2\,d\tau \\ \nonumber
	& = f^2 \Gamma \int_0^\infty |G(u)|^2\,du = \frac{\Gamma}{\Gamma+\Gamma_\perp},
\end{align}
in which the causality $G(u<0)=0$ is used. The norm is independent of $\kappa$.  The optimal pulse shape, obtained by substituting the explicit form of $G(u)$ into Eq. \eqref{eq:xiopt0}, is
\begin{align}\label{xi_opt}
	&\xi_{\mathrm{opt}}(\tau)= \\ \nonumber
	&\begin{cases}
		\displaystyle \sqrt{2\gamma}\,f\,
		e^{-\tfrac{\gamma}{2}(t-\tau)}\frac{\sin\!\big(\Omega (t-\tau)\big)}{\Omega}\,\Theta(t-\tau),
		& \kappa<0, \\[12pt]
		\displaystyle \sqrt{2\gamma}\,f\,e^{-\tfrac{\gamma}{2}(t-\tau)}\,(t-\tau)\,\Theta(t-\tau),
		& \kappa=0, \\[12pt]
		\displaystyle \sqrt{2\gamma}\,f\,e^{-\tfrac{\gamma}{2}(t-\tau)} 
		\frac{2\sinh\!\Big(\tfrac{\sqrt{\kappa}}{2}(t-\tau)\Big)}{\sqrt{\kappa}}\,\Theta(t-\tau),
		& \kappa>0.
	\end{cases}
\end{align}

The time parameter of "$t$" is the time at which the excitation amplitude is evaluated. We can eliminate the arbitrary observation time $t$ by shifting the time origin so that the pulse ends on $t=0$. The optimal single-photon pulse is supported only for negative times, with an effective temporal width that depends on the decay rate $1/\gamma$. The resulting expression for the optimal pulse is
\begin{align}\label{xi_opt1}
	&\xi_{\mathrm{opt}}(\tau)= \\ \nonumber
	&\begin{cases}
		\displaystyle
		-\sqrt{2\gamma}\,f\,
		e^{\tfrac{\gamma}{2}\tau}\,
		\frac{\sin\!\big(\Omega \tau\big)}{\Omega}\,\Theta(-\tau),
		& \kappa<0, \\[12pt]
		\displaystyle
		-\sqrt{2\gamma}\,f\,
		e^{\tfrac{\gamma}{2}\tau}\,\tau\,\Theta(-\tau),
		& \kappa=0, \\[12pt]
		\displaystyle
		-\sqrt{2\gamma}\,f\,
		e^{\tfrac{\gamma}{2}\tau}\,
		\frac{2}{\sqrt{\kappa}}\,
		\sinh\!\Big(\tfrac{\sqrt{\kappa}}{2}\tau\Big)\,\Theta(-\tau),
		& \kappa>0 .
	\end{cases}
\end{align}

We note that the optimal pulse obtained in Eq.~\eqref{xi_opt1} formally extends to $t \to -\infty$, where its exponentially increasing tail makes it non-physical. This pulse, therefore, represents a theoretical bound rather than a physically realizable waveform; in practice, only a finite temporal segment of width $\sim 1/\gamma$ carries significant amplitude and can be implemented experimentally as a truncated version consistent with shaped single-photon envelopes~\cite{Karpinski2017,Sosnicki2023}.

\begin{figure}
	\centering
	\includegraphics[width=0.9\linewidth]{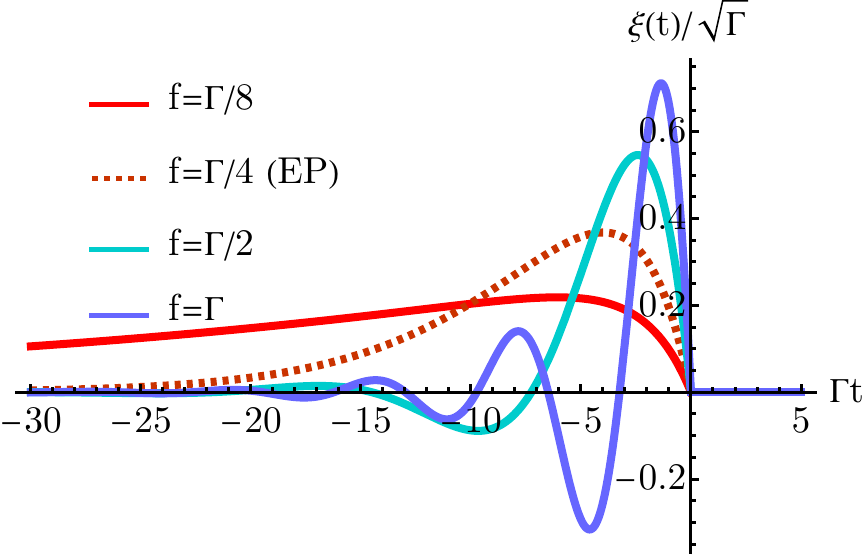}
	\caption{Optimal pulse shapes $\xi_{\mathrm{opt}}(\tau)$ for different values of $f/\Gamma$, plotted as a function of time. The coupling of the atom (charger) to the environment is set to $\Gamma_\perp = 0$.}
	\label{fig:8}
\end{figure}

The shape of the optimal pulse is illustrated in Fig. \ref{fig:8}.

It is useful to compare our analytic pulse optimization with the more common quantum optimal control (QOC) one~\cite{Brif2010,Wang2011}. 
The QOC obtains the control field by numerically optimizing a chosen figure of merit; numerical solutions typically depend on the specific algorithm, initial guess, and convergence criteria.
In this work, the optimized pulse is matched to the time-reversed causal response of the system and can be used as an analytic benchmark for numerical QOC methods for more complex architectures.


\begin{figure}[htbp]
	\centering
	\includegraphics[width=0.9\linewidth]{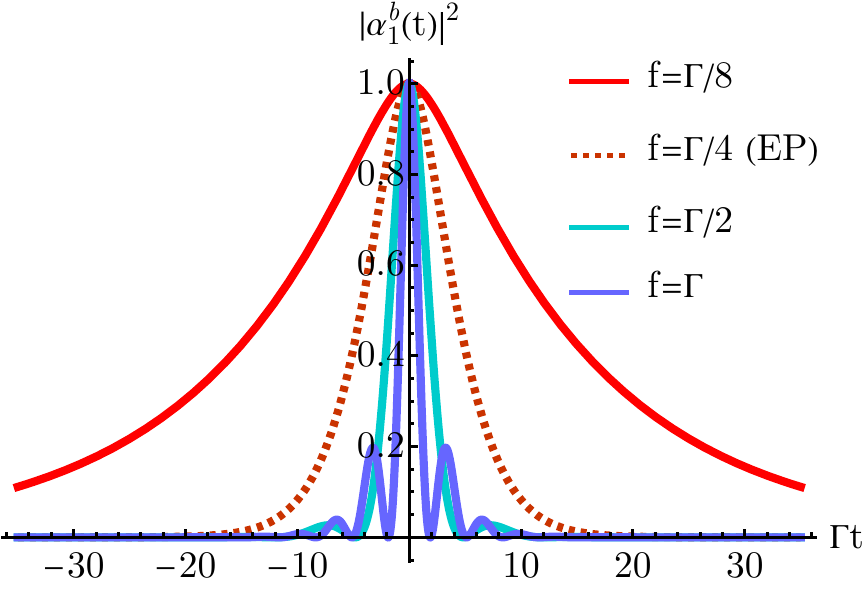}
	\caption{Time evolution and key features of $|\alpha_1^b(t)|^2$ for a optimal input pulse ($\Gamma_\perp = 0$ i.e. $f_{EP}=\Gamma/4$).}
	\label{fig:7}
\end{figure}

The highest excitation probability happens at $t=0$ and is
\begin{equation}\label{eq:alpha_opt}
	|\alpha_1^b(0)|^2 = \frac{\Gamma}{\Gamma+\Gamma_\perp}.
\end{equation}
Analytically obtained $\alpha_1^b$ for different values of $\kappa$ can be written as

\begin{align}
	&\alpha_1^b(t)= -f\sqrt{\Gamma}\\ \nonumber
	& 
	\begin{cases}
		\displaystyle
		\frac{\sqrt{2}\, e^{-\tfrac{1}{2}\gamma t}\, f}{\Omega \,\sqrt{\gamma}\,(\gamma^{2}+4\Omega^{2})} \\
		\Big[
		e^{\gamma t}\,\Theta(-t)\,(\gamma \sin(\Omega t) - 2\Omega \cos(\Omega t)) \\
		\quad+ (\Theta(-t)-1)\,(2\Omega \cos(\Omega t)+\gamma \sin(\Omega t))
		\Big],
		& \kappa<0, \\[1.2em]
		\displaystyle
		\frac{\sqrt{2}\, e^{-\tfrac{1}{2}\gamma t}\, f}{\gamma^{5/2}}
		\Big[
		-\gamma t + \big(e^{\gamma t}(\gamma t - 2) \\ 
        \quad+\gamma t + 2\big)\Theta(-t) - 2
		\Big],
		& \kappa=0, \\[1.2em]
		\displaystyle
		-\frac{2\, e^{-\tfrac{1}{2}(\gamma+\sqrt{\kappa}) t}\, f\,\sqrt{\tfrac{\gamma}{\kappa}}}{\sqrt{2}\,(\gamma^{3}-\gamma \kappa)}
		\Big[
		(\! -1+e^{\sqrt{\kappa} t})\gamma\\ 
        \quad+ (1+e^{\sqrt{\kappa} t})\sqrt{\kappa}+ \Theta(-t) 
		\Big(( -1+e^{\gamma t}) \\ 
        \quad(1+e^{\sqrt{\kappa} t})\sqrt{\kappa}
		-(1+e^{\gamma t})(-1+e^{\sqrt{\kappa} t})\gamma \Big)
		\Big],
		& \kappa>0.
	\end{cases}
\end{align}
and are plotted in Fig. \ref{fig:7} for different values of $f$ and $\Gamma_\perp$ . As can be seen from the figure, the maximum of $|\alpha_1^b|^2$ occurs at $t=0$, and its value coincides with that obtained from Eq. \eqref{eq:alpha_opt}.

\subsection{Minimum-Time Charging}\label{sec:min_time}

Rapid battery charging is essential, which entails minimizing the time required to reach an energy threshold denoted by $p_{\mathrm{th}}$.
 This problem can be solved using the optimal pulse shape introduced in Sec.~\ref {sec:opt_pulse}. We consider the optimal pulse up to a finite time interval $[-\mathcal{T},0]$, with $\mathcal{T}$ being the total pulse duration. 
 The population of the battery's excited state at the end of the pulse is
\begin{equation}
	p(\mathcal{T}) = |\alpha_1^b(0)|^2 = f^2 \Gamma \int_0^\mathcal{T} |G(u)|^2 \, du,
	\label{eq:p_of_T}
\end{equation}
where $G(u)$ is the Green's function defined in Eq.~(\ref{eq:GreenFunction})(cf. Appendix \ref{App:A}). The function $p(\mathcal{T})$ is monotonic with respect to $\mathcal{T}$, and for $\mathcal{T} \to \infty$ it will be saturated to the fundamental upper bound of
\begin{equation}
	\lim_{\mathcal{T} \to \infty} p(\mathcal{T}) = \frac{\Gamma}{\Gamma + \Gamma_\perp}.
\end{equation}
The minimum time for charging can be obtained by solving
\begin{equation}
	p(\mathcal{T}_{\min}) = p_{\rm th}.
\end{equation}
The optimal pulse maximizes the overlap with the system's causal response on each finite time interval. Thus, there is no other pulse shape that can achieve the same target probability $p_{\rm th}$ in a shorter time.
	
At the exceptional point ($\kappa = 0$, i.e.\ $f = \gamma/2$) the Green's function is $G(u) = u e^{-\gamma u/2}$, thus $p(\mathcal{T})$ (from Eq.~\eqref{eq:p_of_T}) is evaluated in closed form of
\begin{equation}
	p(\mathcal{T})=\frac{\Gamma}{\Gamma+\Gamma_\perp} \left[ 1 - e^{-\gamma \mathcal{T}}
	\left( 1 + \gamma \mathcal{T} + \frac{(\gamma \mathcal{T})^2}{2} \right) \right],
\end{equation}
and the minimum charging time follows from the transcendental equation
\begin{equation}\label{eq:transcendental}
	e^{-\gamma \mathcal{T}_{\min}}\left( 1 + \gamma \mathcal{T}_{\min} + \frac{(\gamma \mathcal{T}_{\min})^2}{2} \right)
	= 1 - \frac{p_{\rm th}(\Gamma+\Gamma_\perp)}{\Gamma}.
\end{equation}

Minimizing the time needed to obtain a charging threshold $p_{\mathrm{th}}$ is related to the quantum speed limit (QSL) concept~\cite{Giovannetti2003,Deffner2013}. The QSL discusses lower bounds on the time required to perform a given state transformation under a specified Hamiltonian, and it is typically compared to the results obtained from numerical optimization. In our work, both the optimal protocol and the corresponding minimal time appear together. The truncated optimal pulse maximizes the overlap with the system's causal response on every finite time interval; that is, no other single-photon pulse can reach the same $p_{\mathrm{th}}$ faster. Thus, $T_{\min}$ can be interpreted as the QSL for single-photon charging in this specified quantum battery architecture.

For large probability thresholds that are very close to the fundamental bound of $p_{\rm th}$,  Eq.~(\ref{eq:transcendental}) can be solved using a logarithmic asymptotic approximation, which gives 
\begin{equation}
	\mathcal{T}_{\min} \sim \frac{1}{\gamma} \left[ \ln \left(\frac{1}{1 - q}\right) + \mathcal{O}\left(\ln\ln\left(\frac{1}{1 - q}\right) \right) \right],
\end{equation}
where $q = {p_{\rm th}(\Gamma+\Gamma_\perp)}/{\Gamma}$. This demonstrates that the fundamental charging speed depends on both decay rates of $\Gamma$ and $\Gamma_\perp$. 
where $q = {p_{\rm th}(\Gamma+\Gamma_\perp)}/{\Gamma}$.

Thus, the fastest way to obtain a desirable energy is to use the truncated optimal pulse with duration $\mathcal{T}_{\min}$. Any other pulse shape achieves this threshold strictly more slowly.
In the ideal limit of $\Gamma_\perp=0$, the fundamental bound is $p(T\to\infty)\to 1$, and the time required to reach full charging $p_{\mathrm{th}}=1$ diverges. This limit of population is approached only asymptotically; all experimentally meaningful thresholds $p_{\mathrm{th}}<1$ will be achieved at a finite $T_{\min}$.


\subsection{Power-optimal pulse}
	
	Before discussing the fully power-optimal pulse, we clarify that if a fixed charging threshold $p_{\rm th}$ is desired, optimal-power can be obtained from the minimum-time analysis in Sec.~\ref{sec:min_time}. 
	The truncated optimal pulse leads to the threshold of $p_{\rm th}$ in the shortest possible time of $\mathcal{T}_{\min}$. The corresponding power $P_{\rm th}={p_{\rm th}}/{\mathcal{T}_{\min}}$ is undoubtedly maximal among all pulses achieving the same threshold, i.e., for a fixed stored energy, the minimum-time pulse is in fact
	the optimal-power pulse.
	
	We now discuss a more general question: how the power of $P(T)=p(T)/T$ can be maximized 
	without fixing the charging threshold. Considering the normalized single-photon pulse, for any duration $T$, the pulse shape that maximizes $p(T)$ is the time-reversed, 
	truncated optimal pulse proportional to the system’s causal Green's function of $\xi_{\mathrm{opt}}(t)\propto G(-t)$ ($t\in[-T,0]$). The pulse duration $T$ is the whole time duration over which the pulse exists.    
	The power-optimal solution has the same pulse shape and selects the duration
	\begin{equation}
		T^\ast=\arg\max_{T>0}\frac{p(T)}{T},
	\end{equation}
	which is characterized by the optimization condition
	\begin{equation}
		p'(T^\ast)\,T^\ast=p(T^\ast),
		\qquad 
		p'(T)=f^2\Gamma\,|G(T)|^2.
	\end{equation}
	
	In exceptional point ($\kappa=0$, $f=\gamma/2$), the Green's function is represented by $G(u)=u\,e^{-\gamma u/2}$ and the population of the excited-state is
	\begin{equation}
		p(T)=\frac{\Gamma}{\Gamma+\Gamma_\perp}
		\left[
		1-e^{-\gamma T}
		\left(
		1+\gamma T+\frac{(\gamma T)^2}{2}
		\right)
		\right].
	\end{equation}
	The optimal duration $T^\ast$  satisfies, for $x=\gamma T^\ast$,
	\begin{equation}
		e^{x}=1+x+\frac{x^2}{2}+\frac{x^3}{2},
	\end{equation}
	whose non-trivial solution is $x\simeq 3.389$.  
	Consequently,
	\begin{equation}
		\frac{p(T^\ast)}{\Gamma/(\Gamma+\Gamma_\perp)}\simeq 0.657,
		\qquad
		P^\ast\simeq 0.194\,\gamma\,\frac{\Gamma}{\Gamma+\Gamma_\perp}.
	\end{equation}
	
	The power-optimal solution is the truncated optimal pulse with a duration of 
	$T^\ast$ and there is no other pulse shape that can achieve a larger ratio $p(T)/T$.

\section{Conclusion and Outlook}\label{sec:conclusion}

We have presented a model for charging a quantum harmonic oscillator by a single-photon pulse and a TLS as the charger. We present the dynamics of the quantum battery in different regimes identified by the interplay between coherent interaction and dissipation. 
We present an optimal pulse shape that maximizes the energy stored in the battery and saturates the fundamental energy bound of $\Gamma/(\Gamma+\Gamma_\perp)$.

We have investigated the minimum time required to charge the battery and presented a closed-form expression for the charging probability as a function of the pulse duration, $p(\mathcal{T})$. This allows us to determine the minimal time $\mathcal{T}_{\min}$ required to reach any chosen charging threshold $p_{\rm th}$.
This problem is solved analytically at the exceptional point, which corresponds to the most efficient parameter set for the battery, as it charges faster, and the excitations exhibit no oscillations.
We demonstrate that the fastest charging timescale is determined by the parameters $\Gamma$ and $\Gamma_\perp$. No other pulse shape can achieve a given charging threshold faster, establishing a clear QSL for single-photon charging in these quantum systems. We also study the charging power, showing that the power-optimal protocol is given by a truncated optimal pulse, with a specified duration for which the ratio between stored energy and charging time is maximum.
	
Our model can be applied to current experimental setups that use waveguides and cavity QED, i.e., light interacting with the matter in a controllable manner \cite{Ko2022,Molmer2024,Silberfarb2004,Cook2023,Domokos2002,Konyk2016,Turschmann2019}. 
On the other hand, engineering the temporal shape of single photon pulses can be experimentally done using standard electro-optic or acousto-optic modulators that control the amplitude or phase of the excitation field~\cite{Karpinski2017,Karpinski2021,Sosnicki2023}. Our ongoing project is on the generalization of the present results to the charging of quantum batteries using few-photon quantum light.\\

\begin{acknowledgments}
We thank Dr. Charles Andrew Downing for insightful guidance during the development of this work and for helpful feedback on the manuscript. We also thank Dr. Francesco Campaioli for crucial suggestions on clarifications in the manuscript. 
ED acknowledges the support of the Royal Society via the University Research Fellowship (URF/R1/201158) of Charles Downing (University of Exeter).
\end{acknowledgments}

\bibliographystyle{apsrev4-2}
\bibliography{Refs}
\clearpage
\onecolumngrid
\appendix
\section*{Appendix}
\section{Dynamics of the System} \label{App:A}
In this section we solve the Schr{\"o}dinger equation in the interaction picture as,
\begin{equation}\label{Schrodinger}
	\frac{\partial}{\partial t}\ket{\Psi(t)} = -i H^I(t)\ket{\Psi(t)}
\end{equation}
Using the ansatz (\ref{ansatz2}) in Eq (\ref{Schrodinger}) in near resonance i.e., $\omega_e \simeq \omega_b$ we obtain the coupled equations:
\begin{align}\label{ode1}
	&\frac{d}{dt}\alpha_0^b(t) = - i f  \alpha_1^b(t) - \sqrt{\Gamma}  \chi_{0,P}(t,t) - \sqrt{\Gamma_\perp} \chi_{0,E}(t,t) \\ \label{ode2}
	&\frac{d}{dt}\alpha_1^b(t) = - i f  \, \alpha_0^b(t) \\ \label{ode3}
	&\frac{\partial}{\partial t} \chi_{0,P}(t,\tau)  = \sqrt{\Gamma}  \delta(t-\tau) \, \alpha_0^b(t)  \\ \label{ode4}
	&\frac{\partial}{\partial t} \chi_{0,E}(t,\tau)  = \sqrt{\Gamma_\perp} \, \delta(t-\tau) \alpha_0^b(t) 
\end{align}
Equations \eqref{ode3}-\eqref{ode4} can be solved analytically, thus we have
\begin{align}\label{soln3} 
	&\chi_{0,P}(t,\tau)  = \xi(\tau) + \sqrt{\Gamma}  \theta(t-\tau) \alpha_0^b(\tau) \\ \label{soln4}
	&\chi_{0,E}(t,\tau)  = \sqrt{\Gamma_\perp}  \theta(t-\tau) \alpha_0^b(\tau)
\end{align}
We have now a set of ODE for $\alpha_0^b(t)$ and $\alpha_1^b(t)$,
\begin{align}\label{ode1n}
	&\frac{d}{dt}\alpha_0^b(t) =  -  i f  \alpha_1^b(t)  -  \sqrt{\Gamma} \xi(t) - \frac{\Gamma+\Gamma_\perp}{2} \alpha_0^b(t)  \\ \label{ode2n}
	&\frac{d}{dt}\alpha_1^b(t) = - i f  \, \alpha_0^b(t)
\end{align}
In order to solve the equations of (\ref{ode1n}) and (\ref{ode2n}), we can eliminate $\alpha_1^b(t)$ thus we obtain
\begin{align}\label{ode1nn}
    &\ddot{\alpha}_0^b(t) + \gamma \dot{\alpha}_0^b(t) + f^2  \alpha_0^b(t)  = -\sqrt{\Gamma} \dot{\xi}(t) 
\end{align}
where $\gamma=(\Gamma+\Gamma_\perp)/2$. We define $\kappa = \gamma^2-4f^2$ and,
\begin{equation}
    G(t)=\frac{e^{r_+ t}-e^{r_- t}}{r_+-r_-}, \qquad r_\pm=\frac{-\gamma\pm\sqrt{\kappa}}{2},
\end{equation}
Thus, the general solution can be written as
\begin{equation}
    \alpha_0^b(t)=-\sqrt{\Gamma}\int_{t_0}^t G'(t-\tau)\,\xi(\tau)\,d\tau,
\end{equation}
where,
\begin{align}\label{eq:GreenFunction}
    &G(t) = \frac{2e^{-\gamma t/2}}{\sqrt{\kappa}} \sinh\!\Big(\tfrac{\sqrt{\kappa}}{2}\,t\Big) \\
    &G'(t)=e^{-\gamma t/2}\!\left[\cosh\!\Big(\tfrac{\sqrt{\kappa}}{2}\,t\Big)
    -\frac{\gamma}{\sqrt{\kappa}}\sinh\!\Big(\tfrac{\sqrt{\kappa}}{2}\,t\Big)\right],
\end{align}
Thus for $\alpha_1^b(t)$ we have
\begin{equation} \label{eq:alphag0}
    \alpha_1^b(t)= i f \sqrt{\Gamma}\int_{t_0}^t G(t-\tau)\,\xi(\tau)\,d\tau,
\end{equation}
The sign of $\kappa$ determines the qualitative behavior of the system’s response. 
For $\kappa<0$ (underdamped), the poles are complex conjugates, thus $G(t)$ and $G'(t)$ exhibit oscillations at frequency $\Omega=\sqrt{f^2-\gamma^2/4}$ which are modulated by exponential decay of $e^{-\gamma t/2}$; the response shows resonance-like enhancement near the natural frequency. 
At the EP $\kappa=0$, the poles merge and produce the fastest possible non-oscillatory decay. In this case, $G(t)$ and $G'(t)$ follow purely exponential dynamics with a linear prefactor. This
marks the transition between oscillatory and exponential behavior. 
In $\kappa>0$ (overdamped), the poles are real and negative, thus $G(t)$ and $G'(t)$ decay exponentially, which leads to relaxation without oscillation. 
The value of $\kappa$ governs the balance between coherent coupling $f$ and damping $\gamma$, determining whether the dynamics are oscillatory, at the EP, or purely relaxational. 
We can write $G(t)$ and $G'(t)$ explicitly as
\begin{equation}\label{eq:Green}
    G(t)=
    \begin{cases}
    	{ e^{-\gamma t/2}} \sin(\Omega\,t)/{\Omega}, & \kappa<0,\\[12pt]
    	{ t \, e^{-\gamma t/2}} , & \kappa=0,\\[12pt]
    	{2 \, e^{-\gamma t/2}} \sinh\!\Big(\tfrac{\sqrt{\kappa}}{2}\,t\Big)/{\sqrt{\kappa}}, & \kappa>0.
    \end{cases}
\end{equation}
\begin{equation}\label{eq:Greenprime}
    G'(t)=
    \begin{cases}
    	e^{-\gamma t/2}\!\left[\cos\!\Big(\Omega\,t\Big)
    	-{\gamma}\sin\!\Big(\Omega\,t\Big)/(2\Omega)\right], & \kappa<0,\\[12pt]
    	e^{-\gamma t/2}\!\left[1
    	-{\gamma}t/2\right] , & \kappa=0,\\[12pt]
    	e^{-\gamma t/2}\!\left[\cosh\!\Big(\tfrac{\sqrt{\kappa}}{2}\,t\Big)
    	-{\gamma}\sinh\!\Big(\tfrac{\sqrt{\kappa}}{2}\,t\Big)/{\sqrt{\kappa}}\right], & \kappa>0.
    \end{cases}
\end{equation}
Accordingly, $\alpha_0^b(t)$ and $\alpha_1^b(t)$ are
\begin{equation}
    \alpha_0^b(t)=
    \begin{cases}
    	-\sqrt{\Gamma}\displaystyle\int_{t_0}^t \left[\cos\!\big(\Omega (t-\tau)\big)-\dfrac{\gamma}{2\Omega}\sin\!\big(\Omega (t-\tau)\big)\right]e^{-\gamma (t-\tau)/2}\xi(\tau)\,d\tau, & \kappa<0,\\[12pt]
    	-\sqrt{\Gamma}\displaystyle\int_{t_0}^t \left(1-\dfrac{\gamma (t-\tau)}{2}\right)e^{-\gamma (t-\tau)/2}\,\xi(\tau)\,d\tau, & \kappa=0,\\[12pt]
    	-\sqrt{\Gamma} \displaystyle\int_{t_0}^{t}
    	\left[
    	\cosh\!\Big(\tfrac{\sqrt{\kappa}}{2}(t-\tau)\Big)
    	-\frac{\gamma}{\sqrt{\kappa}}\sinh\!\Big(\tfrac{\sqrt{\kappa}}{2}(t-\tau)\Big)
    	\right]e^{-\gamma (t-\tau)/2}\xi(\tau)\,d\tau, & \kappa>0.
    \end{cases}
\end{equation}
\begin{equation}
    \alpha_1^b(t)=
    \begin{cases}
    	if\frac{\sqrt{\Gamma}}{{\Omega}}\displaystyle\int_{t_0}^t e^{-\gamma (t-\tau)/2} \sin\!\Big(\Omega\,(t-\tau)\Big) \xi(\tau)\,d\tau, & \kappa<0,\\[12pt]
    	if\sqrt{\Gamma}\displaystyle\int_{t_0}^t (t-\tau)e^{-\gamma (t-\tau)/2}\,\xi(\tau)\,d\tau, & \kappa=0,\\[12pt]
    	2if \sqrt{\frac{\Gamma}{\kappa}} \displaystyle\int_{t_0}^{t} e^{-\gamma (t-\tau)/2} \sinh\!\Big(\tfrac{\sqrt{\kappa}}{2}\,(t-\tau)\Big) \xi(\tau)\,d\tau, & \kappa>0.
    \end{cases}
\end{equation}

\section{Different Pulse Shapes} \label{App:C}
\begin{figure}[htbp]
	\centering
	\includegraphics[width=0.4\textwidth]{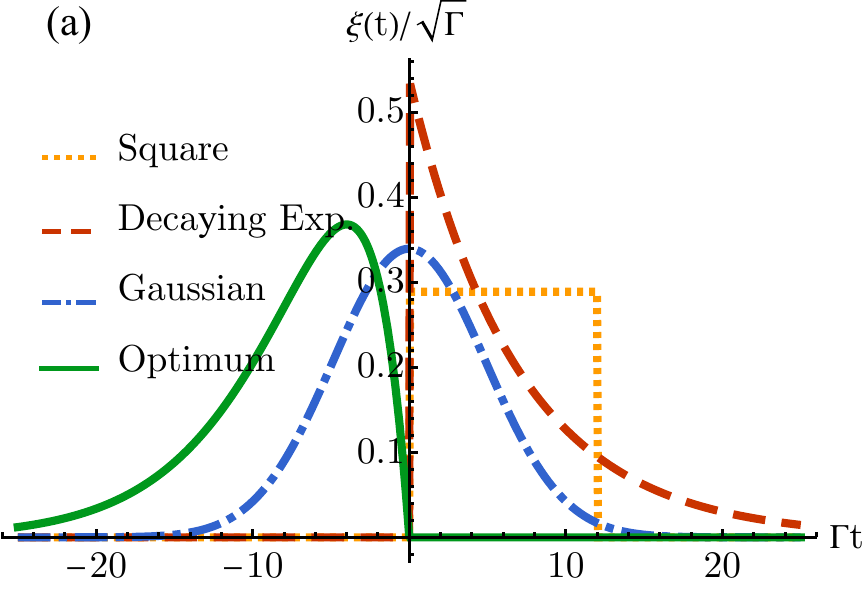}
	\hspace{1cm}
	\includegraphics[width=0.4\textwidth]{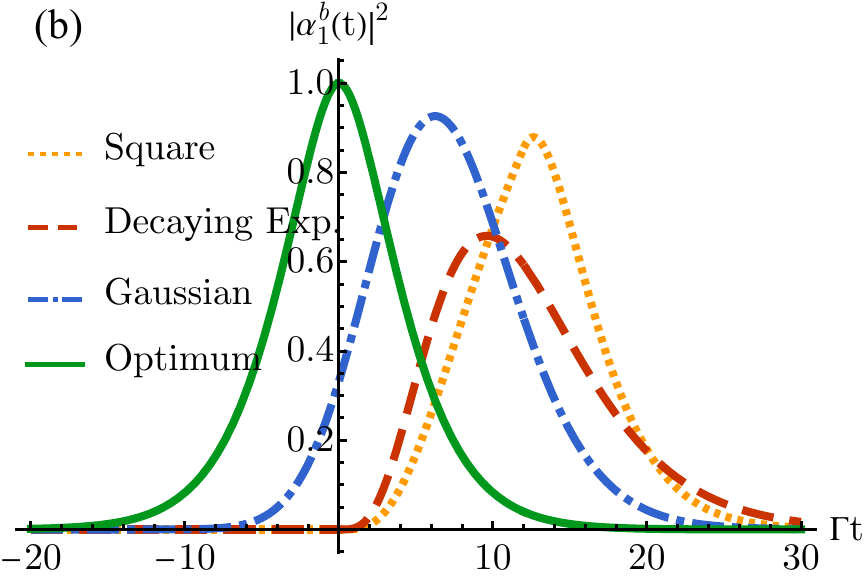}
\caption{ (a) Normalized pulse shapes $\xi(\tau)$ on a timescale set by $\Gamma$, all satisfying $\Gamma T_\sigma = 2\sqrt{3}$. (b) Time-dependent excitation probability $|\alpha_1^b(t)|^2$ at the exceptional point (EP). We assume $\Gamma_\perp = 0$ and optimal coupling $f_{EP} = \Gamma/4$.}
\label{fig:pulseandalpha}
\end{figure}

In this appendix, we present the analytical expressions for the excitation amplitude $\alpha_1^b(t)$ for a few widely used single-photon pulse shapes in quantum optical experiments. The temporal envelopes are listed in Table~\ref{tab:pulseshape}. For a fair comparison, we characterize the pulse duration $T_\sigma$ as the standard deviation of the temporal intensity distribution $\xi(t)^2$. The normalized pulse shapes are shown in Fig.~\ref{fig:pulseandalpha}(a), all chosen to share the same $T_\sigma$.
\begin{table}[htbp]
    \centering
    \caption{Pulse shapes considered in the temporal domain. Here, $\Theta(t)$ denotes the Heaviside step function and $T_\sigma$ is the temporal standard deviation.}
    \begin{tabular}{>{\centering\arraybackslash}p{2.5cm}  >{\centering\arraybackslash}p{4cm}  >{\centering\arraybackslash}p{4cm}}
    \toprule
      Shape  & $\xi(t)$ & $T_\sigma$\\
      \midrule
       Square  & $ \Theta(t)\Theta(T-t)/\sqrt{T}$ & ${T}/{\sqrt{12}}$\\
       Decaying Exp & $ {e^{-{t}/{2T}}}\Theta(t)/{\sqrt{T}}$ & $T$ \\
       Gaussian & $e^{-{t^2}/{4 T^2}}/{(2\pi T^2)^{1/4}}$ & $T$ \\
       Optimum & Eq.~\eqref{xi_opt} & 
       $\begin{cases}
               \sqrt{3}/\gamma &\kappa=0 \\
               \sqrt{\frac{1}{\gamma^2} + \frac{\gamma^2-2f^2}{4f^4}} & \kappa \ne 0
       \end{cases}$\\
       \bottomrule
    \end{tabular}
    \label{tab:pulseshape}
\end{table}
The \textit{square pulse}, given by $\xi(t) = \Theta(t)\Theta(T-t)/\sqrt{T}$, has a duration $T$ and a constant amplitude. Substituting this form into Eq.~\eqref{eq:alphag} leads to
\begin{align}
	&\alpha_1^b(t>0)= \\ \nonumber
	&i\sqrt{\Gamma}\, f\;
	\begin{cases}
		\displaystyle
		\frac{e^{-\tfrac{1}{2}\gamma t}}{\sqrt{T}\,\Omega(\gamma^{2}+4\Omega^{2})}
		\Big[-4\Omega\cos(\Omega t) - 4 e^{\tfrac{T\gamma}{2}}\Omega\cos((t-T)\Omega)(\Theta(T-t)-1) \\
		\quad + 4 e^{\tfrac{\gamma t}{2}}\Omega\,\Theta(T-t) 
		- 2\gamma\sin(\Omega t) - 2 e^{\tfrac{T\gamma}{2}}\gamma(\Theta(T-t)-1)\sin((t-T)\Omega)
		\Big],
		& \kappa<0, \\[1.2em]
		\displaystyle
		-\frac{2 e^{-\tfrac{1}{2}\gamma t}}{\sqrt{T}\,\gamma^{2}}
		\Big[ t\gamma 
		+ e^{\tfrac{T\gamma}{2}}(t\gamma - T\gamma + 2)(\Theta(T-t)-1) - 2 e^{\tfrac{\gamma t}{2}}\Theta(T-t) + 2
		\Big],
		& \kappa=0, \\[1.2em]
		\displaystyle
		\frac{4 e^{-\tfrac{1}{2}\gamma t}}{\sqrt{T}\,\sqrt{\kappa}(\kappa-\gamma^{2})}
		\Big[ \gamma\sinh(\tfrac{t\sqrt{\kappa}}{2})
		- e^{\tfrac{\gamma t}{2}}\sqrt{\kappa}\,\Theta(T-t) \\
		\quad + e^{\tfrac{T\gamma}{2}}\gamma(\Theta(T-t)-1)\sinh(\tfrac{(t-T)\sqrt{\kappa}}{2}) \\
		\quad + \sqrt{\kappa}\cosh(\tfrac{t\sqrt{\kappa}}{2})
		+ e^{\tfrac{T\gamma}{2}}(\Theta(T-t)-1)\sqrt{\kappa}\cosh(\tfrac{(t-T)\sqrt{\kappa}}{2})
		\Big],
		& \kappa>0.
	\end{cases}
\end{align}
The \textit{exponentially decaying pulse} of $\xi(t) = e^{-t/(2T)}\Theta(t)/\sqrt{T}$ with the corresponding excitation amplitude of
\begin{align}
	&\alpha_1^b(t>0)= \\ \nonumber
	& i\sqrt{\Gamma}\, f\;
	\begin{cases}
		\displaystyle
		\frac{2 e^{-\tfrac{1}{2}\gamma t}\sqrt{T}}{\Omega\big((T\gamma-1)^{2}+4T^{2}\Omega^{2}\big)}
		\Big[ 2T\Omega\,e^{\tfrac{t(T\gamma-1)}{2T}} \\
		\quad - 2T\Omega\cos(\Omega t) 
		+ (1-T\gamma)\sin(\Omega t)
		\Big],
		& \kappa<0, \\[1.2em]
		\displaystyle
		\frac{2 e^{-\tfrac{1}{2}\gamma t}\sqrt{T}}{(T\gamma-1)^{2}}
		\Big[ -T\gamma t + t 
		+ 2T\big(e^{\tfrac{t(T\gamma-1)}{2T}}-1\big)
		\Big],
		& \kappa=0, \\[1.2em]
		\displaystyle
		\frac{4 e^{-\tfrac{1}{2}\gamma t}\sqrt{T}}{\sqrt{\kappa}\big((T\gamma-1)^{2}-T^{2}\kappa\big)}
		\Big[ (1-T\gamma)\sinh(\tfrac{t\sqrt{\kappa}}{2})
		- T\sqrt{\kappa}\cosh(\tfrac{t\sqrt{\kappa}}{2}) \\
		\quad + T\sqrt{\kappa}\,e^{\tfrac{t(T\gamma-1)}{2T}}
		\Big],
		& \kappa>0.
	\end{cases}
\end{align}
The \textit{Gaussian pulse}, $\xi(t) = e^{-t^2/(4T^2)}/(2\pi T^2)^{1/4}$ has a symmetric temporal envelope that is ubiquitous in quantum photonic experiments. The related analytic solution for $\alpha_1^b(t)$ is
\begin{align}
	&\alpha_1^b(t)= \\ \nonumber
	& i\sqrt{\Gamma}\, f \;
	\begin{cases}
		\displaystyle
		-\frac{i \, \exp\!\left(\tfrac{1}{4}\big((\gamma -2 i \Omega )^2 T^2 -2 (\gamma +2 i \Omega ) t\big)\right)
		\sqrt[4]{\tfrac{\pi}{2}} \sqrt{T}}
		{2\Omega}
		\\
		\quad\times\Big[
		e^{2 i \Omega t}\!\left(\erf\!\left(\tfrac{t}{2T}+iT\Omega -\tfrac{\gamma T}{2}\right)+1\right)
		+ e^{2 i \gamma \Omega T^2}\!\left(\text{erfc}\!\left(\tfrac{t-(\gamma+2 i \Omega)T^2}{2T}\right)-2\right)
		\Big],
		& \kappa<0, \\[1.2em]
		\displaystyle
		\frac{e^{-\tfrac{t^2}{4T^2}} \sqrt{T}}{\sqrt[4]{2\pi}}
		\Big[
		2T + e^{\tfrac{(t-\gamma T^2)^2}{4T^2}}(t-\gamma T^2)
		\big(\erf\!\left(\tfrac{t-\gamma T^2}{2T}\right)+1\big)\sqrt{\pi}
		\Big],
		& \kappa=0, \\[1.2em]
		\displaystyle
		\frac{2\, e^{\tfrac{1}{4}(\gamma+\sqrt{\kappa})\big((\gamma+\sqrt{\kappa})T^2 - 2t\big)}
		\sqrt[4]{\tfrac{\pi}{2}} \sqrt{T}}
		{2\sqrt{\kappa}}
		\\
		\quad\times\Big[
		e^{\sqrt{\kappa}(t-\gamma T^2)}\!\left(\erf\!\left(\tfrac{(\sqrt{\kappa}-\gamma)T^2+t}{2T}\right)+1\right)
		+ \text{erfc}\!\left(\tfrac{t-(\gamma+\sqrt{\kappa})T^2}{2T}\right) - 2
		\Big],
		& \kappa>0.
	\end{cases}
\end{align}
where $\mathrm{erfc}(x)$ denotes the complementary error function, defined as
\[
\mathrm{erfc}(x) = 1 - \mathrm{erf}(x) = \frac{2}{\sqrt{\pi}} \int_x^{\infty} e^{-t^2}\, dt.
\]
Figure~\ref{fig:pulseandalpha}(b) displays $|\alpha_1^b(t)|^2$ for all pulse shapes under identical charging conditions. This comparison illustrates how temporal engineering of the incident photon affects both the charging efficiency and the maximum stored excitation in the quantum battery.

\section{Charging dynamics for Gaussian pulse shapes}\label{sec:gauss}

\begin{figure}[htbp]
	\centering
	\includegraphics[width=.32\textwidth]{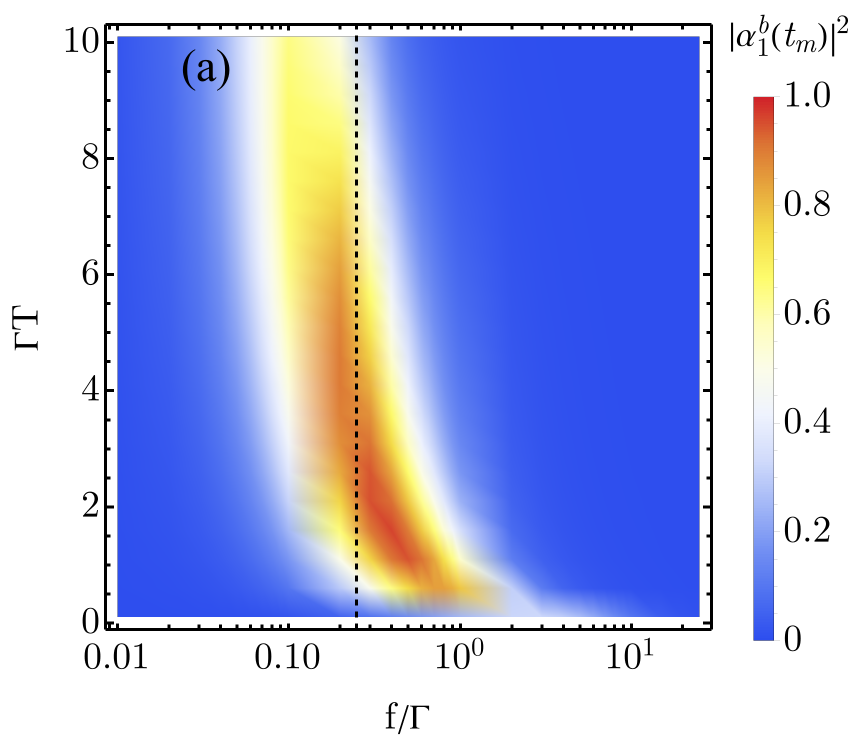}
	\includegraphics[width=.32\textwidth]{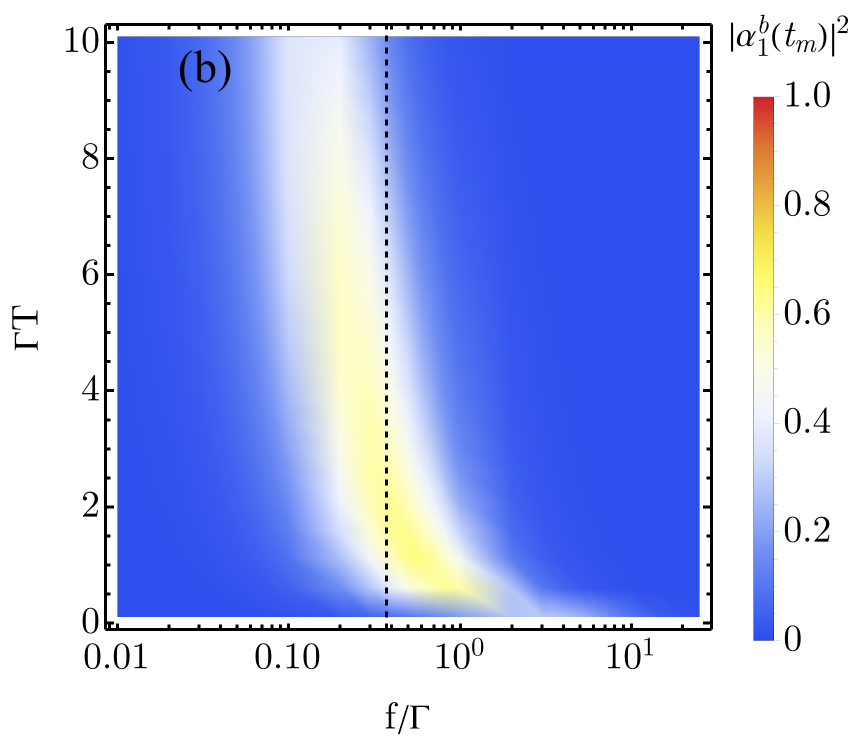}
	\includegraphics[width=.32\textwidth]{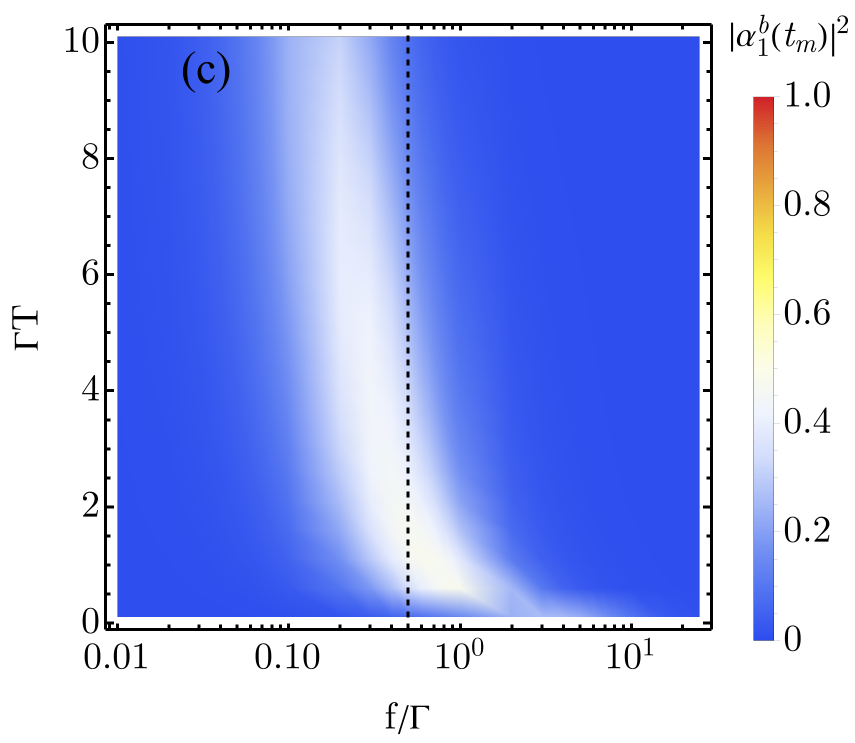}
	\caption{Maximum excitation probability ($|\alpha_1^b(t_m)|^2$) as a function of the dimensionless quantities of $\Gamma T$ and $f/\Gamma$ for a Gaussian input pulse. The decay to the environment is (a) $\Gamma_\perp = 0$ (i.e. $f_{EP}=\Gamma/4$), (b) $\Gamma_\perp = 0.5\Gamma$ (i.e. $f_{EP}=3\Gamma/8$), and (c) $\Gamma_\perp = \Gamma$ (i.e. $f_{EP}=\Gamma/2$) . Vertical dashed lines in panels: guides for the eye at the EP.}
	\label{fig:6}
\end{figure}

\begin{figure}[htbp]
	\centering
	\includegraphics[width=.32\textwidth]{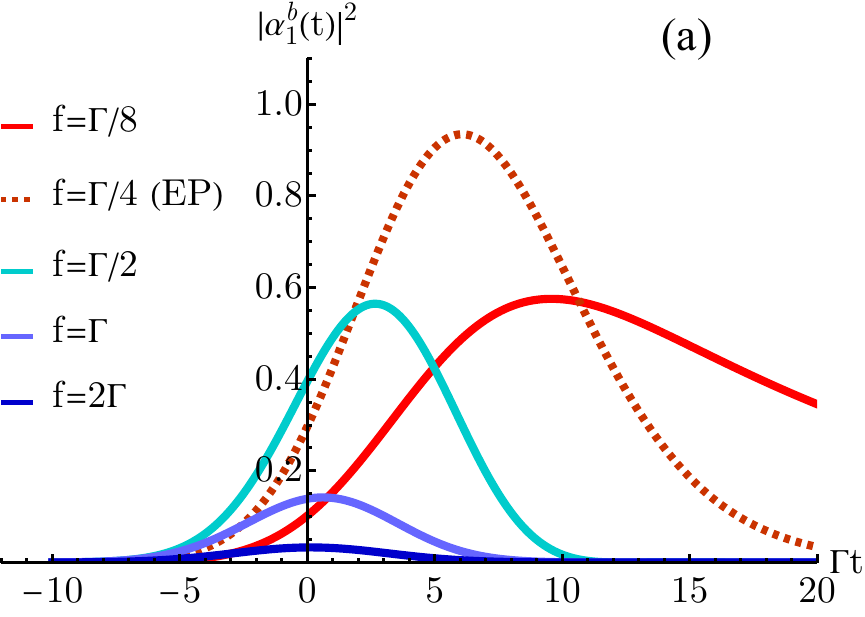}
	\includegraphics[width=.32\textwidth]{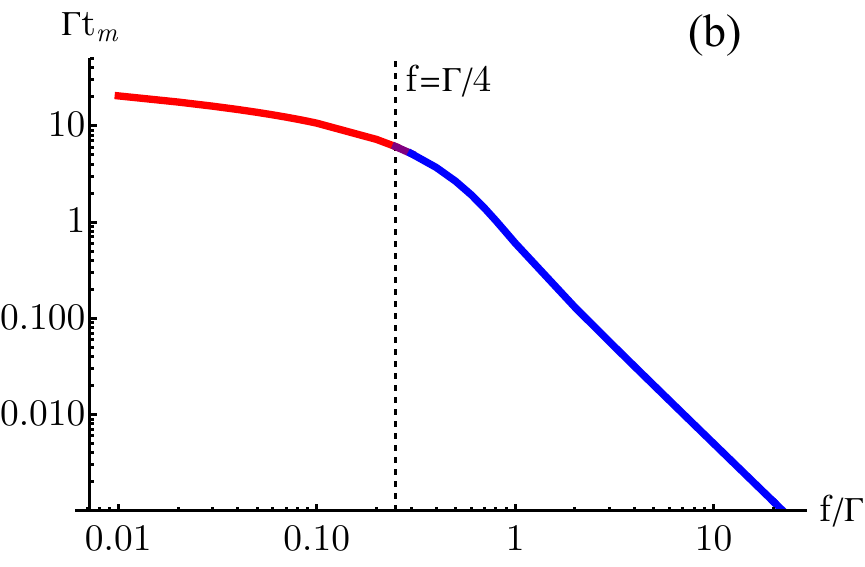}
	\includegraphics[width=.32\textwidth]{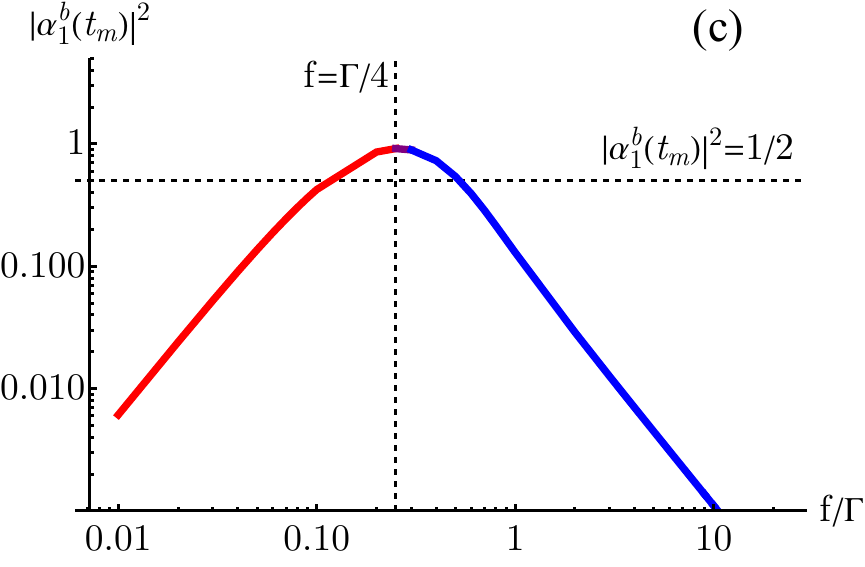}
	\caption{Time evolution and key features of excitation probability of $|\alpha_1^b(t)|^2$ for a Gaussian input pulse. (a) Probability that the HO is excited state; (b) Time $t_m$ at which $|\alpha_1^b(t)|^2$ attains its maximum versus $f/\Gamma$; (c) Maximum of $|\alpha_1^b(t)|^2$ versus $f/\Gamma$. All plots were generated for $\Gamma T = 3$ and $\Gamma_\perp=0$. Vertical dashed lines in panels (b) and (c) are guides for the eye at the EP ($f_{EP}=\Gamma/4$).}
	\label{fig:3n}
\end{figure}

Gaussian pulses are among the most common temporal profiles in single photon pulses. Thus, we analyze their charging performance in more detail in this appendix. The analytical expressions governing the dynamics under Gaussian driving are provided in Appendix~\ref{App:C}; the corresponding plots appear in Figs.~\ref{fig:6}–\ref{fig:3n}.
The finite temporal width of a Gaussian pulse introduces an additional control parameter, $T$, which plays a crucial role in determining the charging efficiency. As shown in Figs.~\ref{fig:6}(a)–(c), increasing the environmental coupling rate $\Gamma_\perp$ reduces the stored excitation due to dissipation. In particular, when $\Gamma_\perp = \Gamma$, the battery remains passive for all pulse duration values.
The time at which the maximum stored energy is achieved varies only weakly with respect to $T$ (not shown here), while the maximum excitation itself depends strongly on $T$ for a fixed $\Gamma_\perp$. In the regime of short pulses, $\Gamma T \approx 0.1$, the interaction time is insufficient to transfer energy efficiently into the battery, leaving it essentially unexcited during the full process (Fig.~\ref{fig:6}(a)).
By contrast, optimal charging occurs at the EP when $\Gamma T \approx 3$, as illustrated in Figs.~\ref{fig:3n}(a)–(c). In this parameter region, the excitation probability reaches its highest attainable value for a Gaussian photon wave packet.
	
\section{\texorpdfstring{Optimal Pulse with $\Gamma_\perp \ne0$}{}}
In this appendix, we present the optimal pulse shape for charging the battery in the presence of an additional electromagnetic environment, which naturally leads to reduced energy storage due to light leakage into the environment.

\begin{figure}[htbp]
		\centering
		\includegraphics[width=0.4\textwidth]{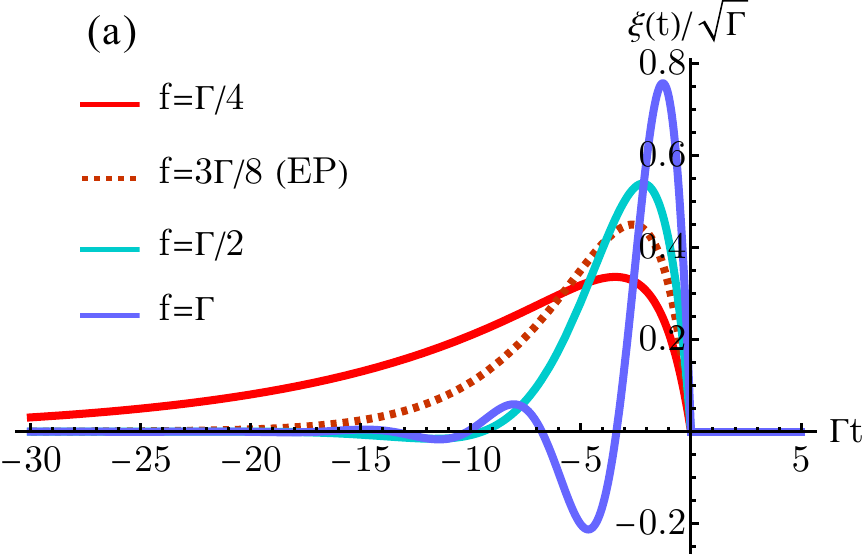}
		\hspace{1cm}
		\includegraphics[width=0.4\textwidth]{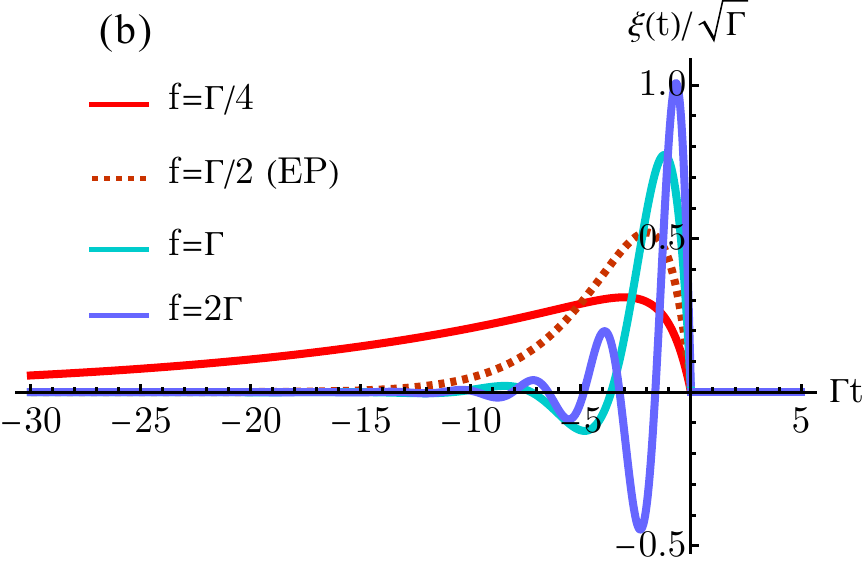}\\
		\includegraphics[width=0.4\textwidth]{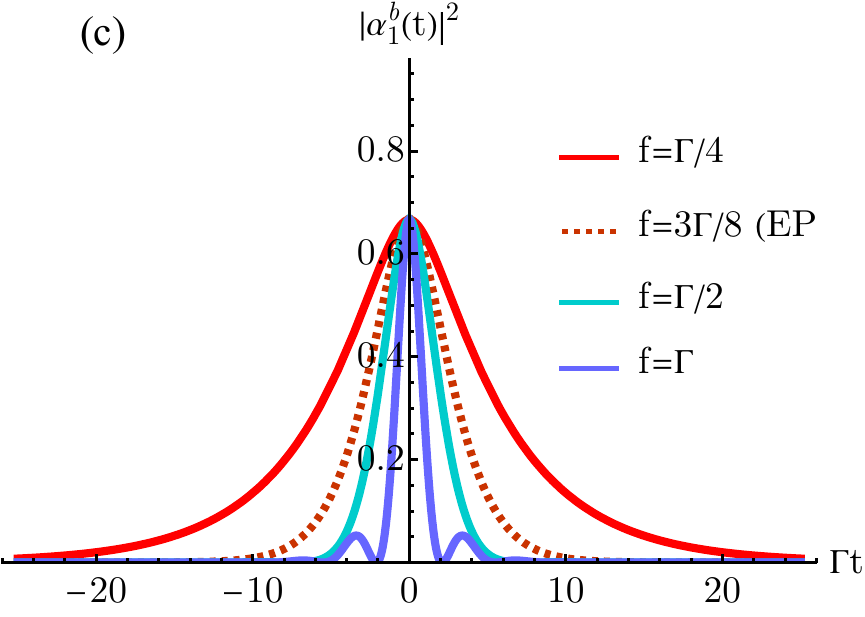}
        \hspace{1cm}
		\includegraphics[width=0.4\textwidth]{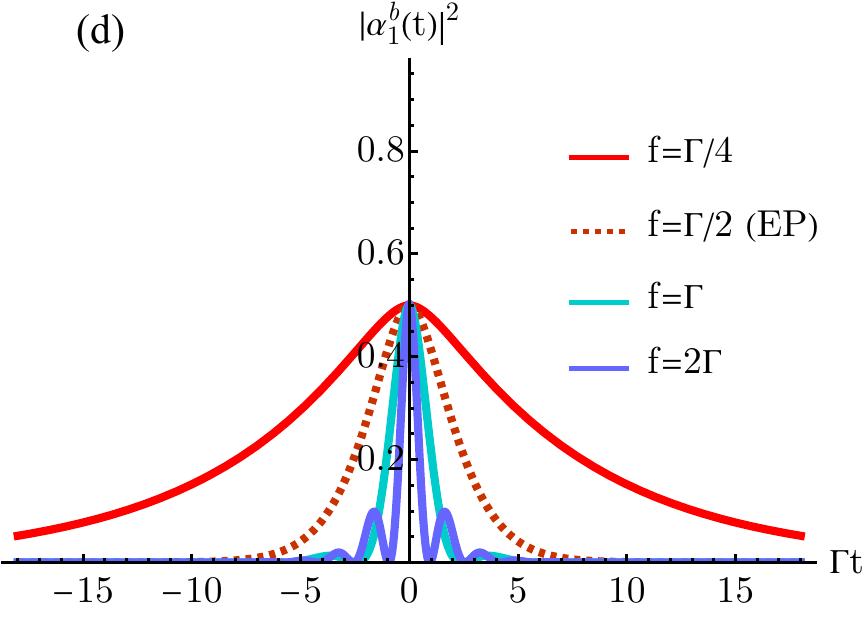}
		\caption{ (a),(b) Optimal pulse shape $\xi_{\mathrm{opt}}(\tau)$ for different values of $f/\Gamma$ as a function of time. (c),(d) time evolution of $|\alpha_1^b(t)|^2$ using the optimal input pulse for charging. The coupling of the atom (charger) to the environment is set to (a),(c) $\Gamma_\perp = 0.5\Gamma$ (i.e. $f_{EP}=3\Gamma/8$), and (b),(d) $\Gamma_\perp = \Gamma$ (i.e. $f_{EP}=\Gamma/2$).}
		\label{fig:7a}
	\end{figure}
\end{document}